\documentclass[sigconf]{acmart}

\usepackage{booktabs} 

\usepackage[utf8]{inputenc}

\title{Do Mobile Developers Ask on Q\&A Sites About Error Codes Thrown by a Cross-Platform App Development Framework? An Empirical Study}

\author{Matias Martinez}
\affiliation{
  \institution{University of Valenciennes, France}
}

\author{Sylvain Lecomte}
\affiliation{
  \institution{University of Valenciennes, France}
}
\usepackage{balance}
\usepackage{numprint}
\usepackage{tabularx} 
\usepackage{mdframed} 
\usepackage{xspace} 
\usepackage{color}
\usepackage{natbib}
\usepackage{graphicx}
\usepackage{verbatim}
\usepackage{listings}
\usepackage{multirow}
\usepackage{framed}
\usepackage{paralist}

\newcommand{\qa}{Q\&A }

\newcommand{\sose}{Stack Overflow}
\newcommand{\so}{Stack Overflow }
\newcommand{\xf}{Xamarin Forum }
\newcommand{\xfse}{Xamarin Forum}
\newcommand{\mtouch}{\emph{Xamarin.iOS} }
\newcommand{\mandroid}{\emph{Xamarin.Android} }
\newcommand{\mtouchse}{\emph{Xamarin.iOS}}
\newcommand{\mandroidse}{\emph{Xamarin.Android}}

\begin{document}

\begin{abstract}
During last years development frameworks have emerged to make easier the development and maintenance of cross-platform mobile applications.
Xamarin framework is one of them: it takes as input an app written in C\# and produces native code for Android, iOS and Windows Mobile platforms.
When using Xamarin, developers can meet \emph{errors}, identified with \emph{codes},  thrown by the framework.
Unfortunately, the Xamarin official documentation does not provide a complete description, solution or workaround for all those codes.
In this paper, we analyze two sites of questions and answers (Q\&A) related to Xamarin for finding questions that mention those error codes.
We found that, in both sites, there are questions written by developers asking about Xamarin errors, and the majority of them have at least one answer.
Our intuition is this discovered information could be useful for giving support to Xamarin developers. 

\end{abstract}

\maketitle

\section{Introduction}

\subsection{Context}

Nowadays, there are billions of smartphone devices around the world which  run either Android or iOS  mobile platforms \cite{mobileshare}. 
A \emph{cross-platform mobile application} is an application that targets more than one mobile platform. For example, Facebook mobile app is cross-platform: it exists one app for Android, another for iOS. 

A traditional approach for developing this kind of apps is to build, for each platform, a \emph{native} application (i.e., an app built to run in a particular mobile platform) using a particular programming language, SDK (Software Development Kit) and toolkits.
Unfortunately, the development of two or more native apps for a cross-platform app increases the costs of development and maintenance \cite{Martinez:2017:TQI}.
Other approaches such as \emph{hybrid-mobile}  have emerged to provide non-native cross-platforms mobile apps \cite{Ali:2016:MCH,Malavolta2015EndUsers,Malavolta2015Hybrid}. However, beyond a good performance for simple apps \cite{Joorabchi2013Challenges,heitkotter2012evaluating}, those apps do not have the same quality that natives for more complex apps, such the Facebook app \cite{Martinez:2017:TQI}.

Xamarin is a development framework by Microsoft which goal is to make easier the development and maintenance of \emph{native} cross-platforms mobile apps by sharing code across all targeted platforms.
Xamarin is \emph{cross-compiler framework} which receives as input an application written in a non-native language  (C\#) and transforms it to  native code for Android, iOS or Windows Phone platform.  

The advantage of using Xamarin framework is that applications for different platforms can be written to share up to 90\% of their code \cite{xamintrodev}.
The framework offers unified API to access common resources across all three platforms, and contains bindings for nearly the entire underlying platform SDKs in both iOS and Android. 

Xamarin offers two main tools: \mtouch (formerly named MonoTouch) and \mandroid for building iOS and Android native apps, respectively. Both tools can be used inside the IDE Visual Studio by Microsoft or by command line.

\subsection{Architecture of Xamarin apps}

When developing cross-platform apps with Xamarin, 
the overall solution structure (which includes all code of the cross-platform app) is organized in a layered architecture that encourages code sharing.
For that, Xamarin proposes three alternative methods for sharing code between cross-platform applications \cite{xamcodesharing}:  
\begin{inparaenum}[\it 1)]
\item Shared Projects, 
\item Portable Class Libraries, and
\item .NET Standard Libraries.
\end{inparaenum}
Let us briefly describe the first one. The \emph{Shared Projects} method groups code into two project types:
\begin{inparaenum}[\it a)]
\item \emph{Shared  project}: 
contains re-usable code to be shared across different platforms,
\item \emph{Platform-specific application projects}: reference and consume the re-usable code and contains platform-specific features, built on components exposed in the Shared project.
Each of them uses either \mandroid or \mtouch to generate the native code.
\end{inparaenum}

\subsection{Problematic}

When generating the native code using either \mtouch or \mandroidse, 
developers can face \emph{errors} thrown by those tools.
For example, a mobile developer reported in the Xamarin Forum\footnote{https://forums.xamarin.com/discussion/92422}
that, after updating the Google Maps package, 
the app compilation threw the error:
\emph{"Error MT5212: Native linking failed, duplicate symbol [..]"}.
Xamarin identifies that error with the code MT5212.

Xamarin documents two catalogs of error codes: 
one with those from \mtouchse, the other with errors from  \mandroidse".
However, those catalogs do not clearly describe  neither the root of the error not the solution for \emph{all} error codes. 
For this reason, developers need to find more information about the code error and its solution in other sources of information, including questions and answers (Q\&A) sites such as \sose.

\vspace{-0.2cm}
\subsection{Research goals}

Our long term goal is to support mobile developers by proposing them extra-information (e.g, solutions, workarounds, augmented error descriptions) when they meet Xamarin error codes. 
In this paper, 
our goal is to know whether two  questions and answers (Q\&A) sites, i.e.,  \so and \xfse, contain additional information for each Xamarin error code.
For that, we first search in both sites for all questions that mention any error code.
We then study whether those questions have at least one answer and whether any of those answers was accepted as correct.
Our intuition is that accepted answers could contain solutions or workarounds for those errors.
The research questions that guide our work are:

\newcommand{\rqnrfreq}{RQ 1: How many  questions mention error codes? }

\newcommand{\rqtopcodes}{RQ 2: Which are the error codes more asked? }

\newcommand{\rqcategoriesquestions}{RQ 3: Which are the  error  categories with more questions?}

\newcommand{\rqcategoriespercentage}{RQ 4: Which are the categories with more proportion of error code asked?}

\newcommand{\rqaccepted}{RQ 4: How many error codes have at least one  {\it a)} answer; {\it b)} accepted answer?}

\rqnrfreq

\rqtopcodes

\rqcategoriesquestions

\rqaccepted

The paper is structured as follows.
Section \ref{sec:codecatalog} presents the two catalogs of error codes extracted from the Xamarin documentation.
Section \ref{sec:methodology} details the research methodology.
Section \ref{sec:evaluation} presents the results.
Section \ref{sec:relatedwork} presents the related works.
Section \ref{sec:conclusion} concludes the paper. 

\section{Error Codes from Xamarin}
\label{sec:codecatalog}

The official documentation of Xamarin includes two catalogs of errors:
one with codes used by \mtouch  \cite{xamioserrors},
the other with codes used by \mandroid \cite{xamabdroiderrors}.
The codes are grouped into \emph{categories}, which are displayed in Table \ref{tab:category_error_description}. 
We now describe each catalog.

\subsection{Error Codes from \mtouch}

Error codes from \mtouch \cite{xamioserrors} are grouped in 9 categories and identified with the prefix MT.
For example, category MT2 groups all error codes related to \emph{"Linker error"}.
Errors code identifiers start with their category identifier followed by a number.
For example, error code  MT2001 \emph{"Could not link assemblies"} is the code 001 of category MT2 "Linker error".
Table \ref{tab:category_error_description} shows the number of error codes per category (column \#Codes). For example, the category MT2 has 21 codes. The  identifiers of codes from a category are not consecutive, i.e., there is a error code MT2102, but none with code MT2100. 
In total, Xamarin documentation reports 360 error codes for \mtouchse.

Error codes descriptions are in general self-explanatory, such as MT8 \emph{"Runtime error"}, even some use Xamarin or Microsoft-related terminology.
For example, 
\emph{mtouch} (aka MonoTouch, now called \mtouchse) tool is the entry point for compiling code for use in iOS devices and to deploy and launch the code on the device.  \emph{MSBuild} (Microsoft Build Engine) is a platform for building applications with is used by Visual Studio. \emph{AOT} (Ahead of Time compilation) compiles code to a native platform.

The information presented in each code error documentation is not uniform and varies across the codes.
There are codes with an error description (e.g., MT2011), others propose a solution or workaround (e.g., MT2016), and others only include a simple error message (e.g, MT2002).

\subsection{Error codes from \mandroid}

Xamarin documentation displays 10 categories for \mandroid errors \cite{xamabdroiderrors}, which are identified with prefix XA.  
Two categories, XA7 and XA8, are described as \emph{"Reserved"} and do not contain any code. In total, Xamarin documentation reports 105 codes for \mandroidse.
Let us to clarify two concepts.
\emph{mandroid} refers to Mono Android, former name of \mandroidse, whereas 
\emph{symlinks}  refers to symbolic link.
The documentation of code errors from \mandroid  is different that one for \mtouch errors: it only includes the code error and one-sentence description. None presents neither a workaround nor a proposed solution.

\begin{table}[t]
\small
\caption{Description of each category of code and number of codes per category (\#Co).}
\begin{tabular}{ll}

\begin{tabular}{|l|l|r|}
\hline
\multicolumn{3}{|c|}{\mtouch}\\
\hline
{Cat.}&  Description & {\#Co} \\
\hline
\hline
MT0 & mtouch& 106  \\ 
\hline
MT1 & Project related& 36  \\ 
\hline
MT2 & Linker& 21 \\ \hline
MT3 & AOT& 9  \\ \hline
MT4 & Code generation& 68\\ \hline
MT5 & GCC \& toolchain&  25 \\ \hline
MT6 & Internal tools& 6\\ \hline
MT7 & MSBuild& 67\\ \hline
MT8 & Runtime& 22\\
\hline
\hline
\multicolumn{2}{|l|}{Total}&360\\
\hline
\multicolumn{3}{c}{}
\end{tabular}
\begin{tabular}{|l|l|r|}
\hline
\multicolumn{3}{|c|}{\mandroid}\\
\hline
{Cat.}&  Description & {\#Co} \\
\hline
\hline

XA0 & mandroids & 17 \\ \hline
XA1 & File copy/symlinks & 20 \\ \hline
XA2 & Linker & 6 \\ \hline
XA3 & AOT&  3\\ \hline
XA4 & Code generations &  23\\ \hline
XA5 & GCC \& toolchains & 17  \\ \hline
XA6 & Internal tools & 3 \\ \hline
XA7 & Reserved & -  \\ \hline
XA8 & Reserved & -  \\ \hline
XA9 & Licensings &  16\\
\hline
\hline
\multicolumn{2}{|l|}{Total}
&105\\
\hline
\end{tabular}

\end{tabular}
\label{tab:category_error_description}
\end{table}

\section{Methodology}
\label{sec:methodology}

In this section we present the methodology used for responding the research questions.

\paragraph{Datasets of Q\&A}
We analyzed Xamarin-related questions provided by two datasets of Xamarin-related questions and answers \cite{1712.09569}: 
\begin{inparaenum}[\it a)]
\item Xamarin Forums, which has 85,908 questions from the Xamarin Forum site; 
\item Xamarin-related \qa extracted from \sose, which has 44,434  questions mined using a technique from Rosen et Shihab \cite{Rosen2016MDA}. 
\end{inparaenum}
In the  remainder  of this paper, when we mention questions from \sose, we refer to those from the latter dataset. We analyzed questions and answers written until September 1st, 2017.

\paragraph{Protocol}
We first extracted all error codes available on the official Xamarin documentation \cite{xamabdroiderrors,xamioserrors}: those iOS-related which have the prefix MT
and those Android-related which have the prefix XA.
We then filtered questions from both datasets which include one or more retrieved error code in:
\begin{inparaenum}[\it a)]
\item question title, or
\item question description.
\end{inparaenum}
Finally, we grouped questions according to two criteria:
\begin{inparaenum}[\it 1)]
\item same code errors; 
\item same error category.
\end{inparaenum}

\section{Evaluation}
\label{sec:evaluation}

\subsection{\rqnrfreq}
\label{sec:rq1}

After filtering questions from \xf and \sose, 
we found  719  and 226 questions, respectively, that mention one or more error code from \mtouchse; 
and 
402 and 104 questions, respectively, that mention one or more error from \mandroidse.

These  questions represent the  1.3\%  (1121 out of 85,908) 
of all questions written in the Xamarin Forum, and the  0.74\% (330 out of 44,434) from all Xamarin-related questions from \sose.

\begin{framed}
{\bf Response RQ 1:}
In \xf and \so \qa sites, there are 1121  and 330 questions, respectively, that mention, at least, one Xamarin error code.
\end{framed}

The 25.3\% (91 out of 360) of the error codes from \mtouch are present on questions from \xfse, whereas in \so the 15.8\% of the errors (57 out of 360) are mentioned.
Regarding with \mandroidse, 40 (38,1\%) and 24 (22.9\%) error codes are in questions from \xf and \sose, respectively.

\vspace{-0.2cm}
\subsection{\rqtopcodes}

\begin{table}[tbp]
\small
\caption{The 10 most mentioned Xamarin error code in \xf (XF) and  \so (SO).}
\begin{tabular}{|l|r|r|r||l|r|r|r|}
\hline
\multirow{2}{*}{Codes}&\multicolumn{3}{c||}{\#Questions} &\multirow{2}{*}{Codes} &\multicolumn{3}{c|}{\#Questions}\\ 
\cline{2-4}
\cline{6-8}
&  {XF} &  {SO} &  {Total} & & XF & SO & Total   \\ 
\hline
MT2002 & 121 & 47 & 168 & XA2006 &  {157} &  {45} &  {202}   \\ 
MT5202 & 70 & 32 & 102 & XA0000 &  {33} &  {11} &  {44}   \\ 
MT3001 & 64 & 20 & 84 & XA5207 &  {32} &  {11} &  {43}   \\ 
MT1006 & 58 & 14 & 72 & XA006 &  {32} &  {2} &  {34}   \\ 
MT5210 & 44 & 16 & 60 & XA5209 &  {20} &  {8} &  {28}   \\ 
MT5211 & 39 & 15 & 54 & XA9005 &  {21} &  {4} &  {25}   \\ 
MT5209 & 34 & 16 & 50 & XA5206 &  {15} &  {6} &  {21}   \\ 
MT2001 & 35 & 2 & 37 & XA9010 &  {17} &  {1} &  {18}   \\ 
MT5201 & 24 & 10 & 34 & XA9006 &  {14} &  {3} &  {17}   \\ 
MT3005 & 22 & 6 & 28 & XA5205 &  {14} &  {2} &  {16}  \\ 
... & ... & ... & ... & ... &  ... &  ... &  ...  \\ 
\hline
Total & 857 & 275 & 1132 &Total  &  {447} &  {117} & 564    \\ 
\hline
|Codes| & 91  &57 &&|Codes|&40 &24 &\\ 
\% &25.3\% &15.8\%&&\%& 38.1\%&22.9\%&\\ 

 \hline
\end{tabular}
\label{tab:topcodes}
\end{table}
 
Table \ref{tab:topcodes} displays the 10 most mentioned error codes in questions from \xf and \sose.
The most mentioned code from \mtouch is MT2002 with 168 questions.
The code description is \emph{"MT2002 Can not resolve reference:*"} and, unlike other error codes from the same category, it does not have neither an associate description (e.g., such that one from MT2011) nor proposed solution (e.g., such that one from MT2016). 
For this reason, it makes sense that developers ask about that error code on \so and \xf  for facing the missing information in the Xamarin documentation.

Moreover, we found that for 4 out of the 10 most mentioned error from \mtouchse, their documentations do not include neither an explanation nor a proposed solution.
Between the other error codes, we observe that
some have a proposed solution: e.g., code MT3001 \emph{"Could not AOT the assembly '*'"} has:  \emph{"disabling incremental builds in the project's iOS Build option"}.

For error coded related to Android,
the most mentioned code is XA2006 \emph{"Reference to metadata item '{0}' [...] could not be resolved"}, and it belongs to error category XA2 \emph{"Linker Errors"}.
The second one, XA0000, does not provide much information: its description is	\emph{Unexpected error - Please fill a bug report at http://bugzilla.xamarin.com}.

Furthermore, the sum of the number of questions per error code (Table \ref{tab:topcodes} row "Total"), is larger that the number of questions reported in section \ref{sec:rq1}. This happens due to questions that mention two or more error codes, e.g.,  question 57081 from \xf  mentions error codes MT5212 and MT5213.\footnote{https://forums.xamarin.com/discussion/57081}

\begin{framed}
{\bf Response RQ 2:}
MT2002 
and XA2006 
are  the most frequent  error codes from \mtouch  and \mandroidse, resp.,  on both \so and \xfse.  

\end{framed}

\subsection{\rqcategoriesquestions}
\begin{table}[tbp]
\small
\caption{For each category $c$, the table shows the number of questions (col. \#Q) that include error codes from $c$, the number of distinct codes from $c$ in questions (col. \#Codes in Q), and the percentage of that number w.r.t all the error codes presented in the Xamarin documentation (col. "\%"). }
\begin{tabular}{|l|r|r|c||r|r|c||r|}
\hline
\multirow{3}{*}{Category} &  \multicolumn{3}{c|}{\xf} &  \multicolumn{3}{c|}{\so}  & \#Codes \\ 
 \cline{2-7}
 &  {\#Q} &  {\#Codes} & \% &  {\#Q} &  {\#Codes} & \% &  {in docs} \\
 &   &  {in Q} &  &   &  {in Q} &  &  {} \\
 
 \hline
MT0 & 109 & 34 & 32.1 \ & 26 & 13 & 12.3 \ & 106 \\ 
MT1 & 126 & 19 & 52.8 \ & 36 & 14 & 38.9 \ & 36 \\ 
MT2 & 159 & 5 & 23.8 \ & 51 & 3 & 14.3 \ & 21 \\ 
MT3 & 83 & 4 & 44.4 \ & 25 & 3 & 33.3 \ & 9 \\ 
MT4 & 37 & 9 & 13.2 \ & 12 & 7 & 10.3 \ & 68 \\ 
MT5 & 202 & 18 & 72.0 \ & 75 & 15 & 60.0 \ & 25 \\ 
MT6 & 3 & 2 & 33.3 \ & 1 & 2 & 33.3 \ & 6 \\ 
MT7/8 & 0 & 0 & - & 0 & 0 & - & 22 \\ 
 \hline
Total & 719 & 91 & 25.3 \ & 226 & 57 & 15.8 \ & 360 \\ 

  \hline
   \hline
XA0 & 92 & 11 & 64.7 \ & 24 & 7 & 41.2 \ & 17 \\ 
XA1 & 0 & 0 & - \ & 0 & 0 & - \ & 20 \\ 
XA2 & 157 & 1 & 16.7 \ & 45 & 1 & 16.7 \ & 6 \\ 
XA3 & 6 & 1 & 33.3 \ & 0 & 0 & 0.0 \ & 3 \\ 
XA4 & 18 & 5 & 21.7 \ & 4 & 2 & 8.7 \ & 23 \\ 
XA5 & 75 & 9 & 52.9 \ & 23 & 8 & 47.1 \ & 17 \\ 
XA6/7/8 & 0 & 0 &  -& 0 & 0 & - & 0 \\ 
XA9 & 54 & 13 & 81.3 \ & 8 & 6 & 37.5 \ & 16 \\ 
 \hline
Total & 402 & 40 & 38.1 \ & 104 & 24 & 22.9 \ & 105 \\ 
 \hline
\end{tabular}
\label{tab:codecategoriesfounds}
\end{table}
 
Table \ref{tab:codecategoriesfounds} displays the number of questions  that mention codes from a given category (columns \#Q) and the total number of  codes from a given category that are mentioned by 1+ questions (columns \#Codes in Q).

The code categories from \mtouch  most mentioned are in both \qa sites:  
MT5 (with 202 and 75 questions, resp. i.e., the 29.3\% of asked questions),
MT2, 
and 
MT1.
Moreover, there are two categories MT7 (\emph{"MSBuild"}) and MT8 (\emph{"Runtime"}) whose codes are not mentioned any question.
The most mentioned code categories from \mandroidse, 
are:
XA2 (with 157 and 45 questions, i.e., the 39.9\% of asked questions), 
XA0, 
and 
XA5.
Here, there are 4 categories whose error codes are never mentioned:
XA1 (\emph{"File copy/symlinks"}), XA6 (\emph{"Internal Tools"}), XA7 and XA8 (both  \emph{"Reserved"}).

\begin{framed}
{\bf Response RQ 3:} The error categories with questions more asked are: 
MT5 (\emph{GCC and toolchaine errors}) with the 29.3\%  of asked questions
and 
XA2 (\emph{Linker errors}) with the 39.9\%.
\end{framed}

Moreover, Table \ref{tab:codecategoriesfounds}   displays the percentage of error codes from a category that are present in 1+ questions (Column \%).
In \mtouch, the category that has a higher percentage is MT5: the 72\% (18 out of 25) of its codes are present in 1+ question  in \xf whereas the 60\% (15 out of 25) in \sose. 
Regarding with codes from \mandroidse,  the 81.3\% of codes from category XA9 have 1+ questions on \sose, whereas in \xf its percentage is lower: 37.6\%.

\vspace{-0.2cm}
\subsection{\rqaccepted}

\begin{table}[t]
\small
\caption{Error codes grouped by \% of answered/accepted answers. 
Each cell displays the number of codes where the questions that mention them have a given percentage (i.e., 0\%, $>$0\% or 100\%) of answered/accepted answers.
}
\begin{tabular}{|r|r|r|r|r|r|}
\hline
\multicolumn{2}{|c|}{}&\multicolumn{2}{c|}{\xf}&\multicolumn{2}{c|}{\so}\\ 
\cline{2-6}
& \% & XA & MT & XA & MT \\ \hline

\multirow{3}{*}{\rotatebox[origin=c]{90}{{\scriptsize Answered}}}& 0\% & {5} (12.50\%) & {7} (7.69\%) & {6} (25\% ) & {4} (7.02\%) \\ 
\cline{2-6}
&$>$0\%& 35 (87.5\%) & 84 (92.3\%) & 18 (75\%)	&53 (92.98\%)\\ 
\cline{2-6}
&100\% & {15} (37.50\%) & {47} (51.65\%) & {10} (41.67\%) & {29} (50.88\%) \\ 
\hline
\hline
\multirow{3}{*}{\rotatebox[origin=c]{90}{Accepted}}&0\%&	26 (65\%)&	56 (61.54\%) &	11 (45.83\%)&	8 (14.04\%)\\
\cline{2-6}
&$>$0\%&	14 (35\%)&	35 (38.46\%)&	13 (54.17\%)&	49 (85.96\%)\\
\cline{2-6}
&100\%&	0 (0\%)&	4 (4.40\%)&	4 (16.67\%)	&7 (12.28\%)\\
\hline
\hline
\multicolumn{2}{|c|}{\#Codes} & {40} & {91} & {24} & {57} \\ \hline

\end{tabular}
\label{tab:reponses_accepted}
\end{table}

 Table  \ref{tab:reponses_accepted} presents the number of error codes mentioned by questions Q that have
\begin{inparaenum} [\it a)]
\item zero answers (0\%), i.e., none question from Q was answered,
\item one or more ($>$0\%), i.e., 1+ question from Q with 1+ answer, or
\item all questions from Q with 1+ answer (100\%).
\end{inparaenum}
The table also shows a similar analysis for questions with \emph{accepted} answers.

\subsubsection{Codes with answered questions}

In \xfse, 
the 87.5\% (35 out of 40)  and  92.3\% (84 out of 91) of error codes from \mtouch and \mandroidse, resp., are mentioned by, at least, one question which was answered.
Moreover, those mentioned by all answered questions (i.e., 100\%) represent the 37.5\% and the 51.6\%, resp.
In \sose, we observe a similar trend for codes \mtouchse. 
However, \mandroid codes  mentioned only by not answered  questions (i.e., 0\%) is proportionally higher: 25\%. 

\subsubsection{Codes with accepted answers}

In \xfse, the 35\% and 38.46\% of codes from \mandroid and \mtouch are mentioned by 1+ questions with 1 accepted answer.
In \sose, those percentages are higher: 54.17\% and 85.96\%, resp.,
which means that, for the majority of the error codes, there is 1+ question with an accepted answers.
Moreover, there are codes  (4 from \mandroid and 11 from \mtouchse) which have an accepted answer for \emph{all} the questions that mention them.
We inspect those questions finding that they have, at most, 2 answers per question.

\begin{framed}
{\bf Response RQ 4:}
{\it a)} The 92.6\% (127 codes) and 82.8\% (53) of error codes from \mtouch and  \mandroidse, respectively, have 1+ answered question.
Moreover, 
{\it b)} the 56.7\% (84 codes) and 42.1\% (27) of error codes from \mtouch and  \mandroidse, respectively, have one accepted answer.
\end{framed}

\section{Related work}
\label{sec:relatedwork}

There are several works that classify, compare and evaluate cross-platform mobile application development tools  to build  hybrid mobile  and native apps \cite{heitkotter2012evaluating, majchrzak2017comprehensive,francese2013supporting,dalmasso2013survey,palmieri2012comparison,desruelle2012challenges, Marinho2015NativeMultiple,Ribeiro2012cross}. 
To the best of our knowledge, only one work \cite{1712.09569} focuses on Xamarin, which studies  \qa sites for discovering the main topics of Xamarin related question.
Other works \cite{Malavolta:2017:AIS, majchrzak2018progressive}  focus on Progressive Web Apps (PWAs), a technology introduced by Google for improving  Web mobile apps.
Regarding with the testing  of cross-platform applications,  
CHECKCAMP \cite{Joorabchi2015inco} is a tool for helping mobile developers to test their apps across multiple platforms.
DIFFDROID \cite{Fazzini:2017:ACI}, a  technique that helps developers automatically find cross-platform inconsistencies.

Other works \cite{hu2016crossconsistency,Noei2017,Martin:2016:CIA} have studied the quality of cross-platforms mobile applications by analyzing apps stores such Google Play. 
For example, comparison of user-perceived ratings of cross-platform app \cite{Ali2017SAD}, and comparison between rankings of  hybrid mobile apps and  native apps \cite{Malavolta2015Hybrid}.
To the best of our knowledge, no work has studied Xamarin apps from the apps stores

Previous works have studied Q\&A sites (e.g., \sose) to mine questions about mobile platforms.
For example, 
studies about: 
questions and activities in \so when changes on Android APIs occur \cite{Linares-Vasquez:2014:ACT},
posts from \so  related to iOS and Android  APIs to find API usage obstacles \cite{Wang:2013:DAU},  
questions about Android permission use on \so \cite{Stevens:2013:APU},  
code reuse on \so from Android apps \cite{Abdalkareem2017Reuse},  
self-explanatory of code fragment present in a thread of \so \cite{Treude2017CF}, 
impact of platform dependence on source code quality \cite{Syer2015}.
 
\vspace{-0.2cm}
\section{Conclusion}
\label{sec:conclusion}

Xamarin is a cross-compiler framework  for  building mobiles apps that target more than one platforms.
In this paper we analyzed two  sites of questions and answers (Q\&A) 
for mining questions that mention code errors thrown by the Xamarin platform.
We found that there are  1121  and 330 questions from \xf and \so \qa sites, resp., that mention one or more Xamarin error code. 
This result shows that developers search for additional information to face those errors.
Moreover, the 87.5\%  and  92.3\%, resp., of error codes found on those questions have 1+ answer which could potentially be useful to clarify the error.

For future work, we plan to analyze answers corresponding to those questions to help developers when they meet such error codes by proposing an augmented explanations about the error, candidate solutions and workarounds.

\bibliographystyle{plain}
\balance
\bibliography{references}

\begin{thebibliography}{10}

\bibitem{xamintrodev}
Introduction to mobile development.
\newblock
  \url{https://developer.xamarin.com/guides/cross-platform/getting\_started/introduction\_to\_mobile\_development/}.
\newblock Accessed: 2018-1-20.

\bibitem{xamcodesharing}
Xamarin sharing code options.
\newblock
  \url{https://developer.xamarin.com/guides/cross-platform/application\_fundamentals/code-sharing/}.
\newblock Accessed: 2018-1-20.

\bibitem{xamabdroiderrors}
Xamarin.android errors.
\newblock
  \url{https://developer.xamarin.com/guides/android/troubleshooting/errors/}.
\newblock Accessed: 2018-1-20.

\bibitem{xamioserrors}
Xamarin.ios errors.
\newblock
  \url{https://developer.xamarin.com/guides/ios/troubleshooting/mtouch-errors/}.
\newblock Accessed: 2018-1-20.

\bibitem{Abdalkareem2017Reuse}
Rabe Abdalkareem, Emad Shihab, and Juergen Rilling.
\newblock On code reuse from stackoverflow: An exploratory study on android
  apps.
\newblock {\em Information and Software Technology}, 88(Supplement C):148 --
  158, 2017.

\bibitem{Ali2017SAD}
Mohamed Ali, Mona~Erfani Joorabchi, and Ali Mesbah.
\newblock Same app, different app stores: A comparative study.
\newblock In {\em Proceedings of the 4th International Conference on Mobile
  Software Engineering and Systems}, MOBILESoft '17, pages 79--90, Piscataway,
  NJ, USA, 2017. IEEE Press.

\bibitem{Ali:2016:MCH}
Mohamed Ali and Ali Mesbah.
\newblock Mining and characterizing hybrid apps.
\newblock In {\em Proceedings of the International Workshop on App Market
  Analytics}, WAMA 2016, pages 50--56, New York, NY, USA, 2016. ACM.

\bibitem{dalmasso2013survey}
Isabelle Dalmasso, Soumya~Kanti Datta, Christian Bonnet, and Navid Nikaein.
\newblock Survey, comparison and evaluation of cross platform mobile
  application development tools.
\newblock In {\em 2013 9th International Wireless Communications and Mobile
  Computing Conference (IWCMC)}, pages 323--328. IEEE, 2013.

\bibitem{desruelle2012challenges}
Heiko Desruelle, John Lyle, Simon Isenberg, and Frank Gielen.
\newblock On the challenges of building a web-based ubiquitous application
  platform.
\newblock In {\em Proceedings of the 2012 ACM Conference on Ubiquitous
  Computing}, pages 733--736. ACM, 2012.

\bibitem{Fazzini:2017:ACI}
Mattia Fazzini and Alessandro Orso.
\newblock Automated cross-platform inconsistency detection for mobile apps.
\newblock In {\em Proceedings of the 32Nd IEEE/ACM International Conference on
  Automated Software Engineering}, ASE 2017, pages 308--318, Piscataway, NJ,
  USA, 2017. IEEE Press.

\bibitem{francese2013supporting}
Rita Francese, Michele Risi, Genoveffa Tortora, and Giuseppe Scanniello.
\newblock Supporting the development of multi-platform mobile applications.
\newblock In {\em 2013 15th IEEE International Symposium on Web Systems
  Evolution (WSE)}, pages 87--90. IEEE, 2013.

\bibitem{heitkotter2012evaluating}
Henning Heitk{\"o}tter, Sebastian Hanschke, and Tim~A Majchrzak.
\newblock Evaluating cross-platform development approaches for mobile
  applications.
\newblock In {\em International Conference on Web Information Systems and
  Technologies}, pages 120--138. Springer, 2012.

\bibitem{hu2016crossconsistency}
Hanyang Hu, Cor-Paul Bezemer, and Ahmed~E Hassan.
\newblock Studying the consistency of star ratings and the complaints in 1 \&
  2-star user reviews for top free cross-platform android and ios apps.
\newblock {\em PeerJ Preprints}, 4:e2589v1, November 2016.

\bibitem{mobileshare}
IDC.
\newblock Smartphone os market share, 2017 q1, 2017.

\bibitem{Joorabchi2015inco}
M.~E. Joorabchi, M.~Ali, and A.~Mesbah.
\newblock Detecting inconsistencies in multi-platform mobile apps.
\newblock In {\em 2015 IEEE 26th International Symposium on Software
  Reliability Engineering (ISSRE)}, pages 450--460, Nov 2015.

\bibitem{Joorabchi2013Challenges}
M.~E. Joorabchi, A.~Mesbah, and P.~Kruchten.
\newblock Real challenges in mobile app development.
\newblock In {\em 2013 ACM / IEEE International Symposium on Empirical Software
  Engineering and Measurement}, pages 15--24, Oct 2013.

\bibitem{Linares-Vasquez:2014:ACT}
Mario Linares-V\'{a}squez, Gabriele Bavota, Massimiliano Di~Penta, Rocco
  Oliveto, and Denys Poshyvanyk.
\newblock How do api changes trigger stack overflow discussions? a study on the
  android sdk.
\newblock In {\em Proceedings of the 22Nd International Conference on Program
  Comprehension}, ICPC 2014, pages 83--94, New York, NY, USA, 2014. ACM.

\bibitem{majchrzak2017comprehensive}
Tim Majchrzak and Tor-Morten Gr{\o}nli.
\newblock Comprehensive analysis of innovative cross-platform app development
  frameworks.
\newblock In {\em Proceedings of the 50th Hawaii International Conference on
  System Sciences}, 2017.

\bibitem{majchrzak2018progressive}
Tim~A Majchrzak, Andreas Bi{\o}rn-Hansen, and Tor-Morten Gr{\o}nli.
\newblock Progressive web apps: the definite approach to cross-platform
  development?
\newblock In {\em Proceedings of the 51st Hawaii International Conference on
  System Sciences}, 2018.

\bibitem{Malavolta2015EndUsers}
I.~Malavolta, S.~Ruberto, T.~Soru, and V.~Terragni.
\newblock End users' perception of hybrid mobile apps in the google play store.
\newblock In {\em 2015 IEEE International Conference on Mobile Services}, pages
  25--32, June 2015.

\bibitem{Malavolta:2017:AIS}
Ivano Malavolta, Giuseppe Procaccianti, Paul Noorland, and Petar
  Vukmirovi\'{c}.
\newblock Assessing the impact of service workers on the energy efficiency of
  progressive web apps.
\newblock In {\em Proceedings of the 4th International Conference on Mobile
  Software Engineering and Systems}, MOBILESoft '17, pages 35--45, Piscataway,
  NJ, USA, 2017. IEEE Press.

\bibitem{Malavolta2015Hybrid}
Ivano Malavolta, Stefano Ruberto, Tommaso Soru, and Valerio Terragni.
\newblock Hybrid mobile apps in the google play store: An exploratory
  investigation.
\newblock In {\em Proceedings of the Second ACM International Conference on
  Mobile Software Engineering and Systems}, MOBILESoft '15, pages 56--59,
  Piscataway, NJ, USA, 2015. IEEE Press.

\bibitem{Marinho2015NativeMultiple}
Euler~Horta Marinho and Rodolfo~Ferreira Resende.
\newblock Native and multiple targeted mobile applications.
\newblock In Osvaldo Gervasi, Beniamino Murgante, Sanjay Misra, Marina~L.
  Gavrilova, Ana Maria Alves~Coutinho Rocha, Carmelo Torre, David Taniar, and
  Bernady~O. Apduhan, editors, {\em Computational Science and Its Applications
  -- ICCSA 2015}, pages 544--558, Cham, 2015. Springer International
  Publishing.

\bibitem{Martin:2016:CIA}
William Martin, Federica Sarro, and Mark Harman.
\newblock Causal impact analysis for app releases in google play.
\newblock In {\em Proceedings of the 2016 24th ACM SIGSOFT International
  Symposium on Foundations of Software Engineering}, FSE 2016, pages 435--446,
  New York, NY, USA, 2016. ACM.

\bibitem{1712.09569}
Matias Martinez and Sylvain Lecomte.
\newblock Discovering discussion topics about development of cross-platform
  mobile applications using a cross-compiler development framework, 2017.

\bibitem{Martinez:2017:TQI}
Matias Martinez and Sylvain Lecomte.
\newblock Towards the quality improvement of cross-platform mobile
  applications.
\newblock In {\em Proceedings of the 4th International Conference on Mobile
  Software Engineering and Systems}, MOBILESoft '17, pages 184--188,
  Piscataway, NJ, USA, 2017. IEEE Press.

\bibitem{Noei2017}
Ehsan Noei, Mark~D. Syer, Ying Zou, Ahmed~E. Hassan, and Iman Keivanloo.
\newblock A study of the relation of mobile device attributes with the
  user-perceived quality of android apps.
\newblock {\em Empirical Software Engineering}, 22(6):3088--3116, Dec 2017.

\bibitem{palmieri2012comparison}
Manuel Palmieri, Inderjeet Singh, and Antonio Cicchetti.
\newblock Comparison of cross-platform mobile development tools.
\newblock In {\em Intelligence in Next Generation Networks (ICIN), 2012 16th
  International Conference on}, pages 179--186. IEEE, 2012.

\bibitem{Ribeiro2012cross}
A.~Ribeiro and A.~R. da~Silva.
\newblock Survey on cross-platforms and languages for mobile apps.
\newblock In {\em 2012 Eighth International Conference on the Quality of
  Information and Communications Technology}, pages 255--260, Sept 2012.

\bibitem{Rosen2016MDA}
Christoffer Rosen and Emad Shihab.
\newblock What are mobile developers asking about? a large scale study using
  stack overflow.
\newblock {\em Empirical Softw. Engg.}, 21(3):1192--1223, June 2016.

\bibitem{Stevens:2013:APU}
Ryan Stevens, Jonathan Ganz, Vladimir Filkov, Premkumar Devanbu, and Hao Chen.
\newblock Asking for (and about) permissions used by android apps.
\newblock In {\em Proceedings of the 10th Working Conference on Mining Software
  Repositories}, MSR '13, pages 31--40, Piscataway, NJ, USA, 2013. IEEE Press.

\bibitem{Syer2015}
Mark~D. Syer, Meiyappan Nagappan, Bram Adams, and Ahmed~E. Hassan.
\newblock Studying the relationship between source code quality and mobile
  platform dependence.
\newblock {\em Software Quality Journal}, 23(3):485--508, Sep 2015.

\bibitem{Treude2017CF}
C.~Treude and M.~P. Robillard.
\newblock Understanding stack overflow code fragments.
\newblock In {\em 2017 IEEE International Conference on Software Maintenance
  and Evolution (ICSME)}, pages 509--513, Sept 2017.

\bibitem{Wang:2013:DAU}
Wei Wang and Michael~W. Godfrey.
\newblock Detecting api usage obstacles: A study of ios and android developer
  questions.
\newblock In {\em Proceedings of the 10th Working Conference on Mining Software
  Repositories}, MSR '13, pages 61--64, Piscataway, NJ, USA, 2013. IEEE Press.

\end{thebibliography}
\end{document}